\documentclass[%
reprint,
 amsmath,amssymb,
 aps,
 prx,
]{revtex4-2}

\usepackage{graphicx}
\usepackage{dcolumn}
\usepackage{bm}
\usepackage{xcolor}
\usepackage{cancel}
\usepackage{ dsfont }
\usepackage{booktabs}
\usepackage{comment}
\usepackage[T1]{fontenc}

\newcommand*{\bra}{\langle}
\newcommand*{\ket}{\rangle}
\newcommand*{\ua}{\uparrow}
\newcommand*{\da}{\downarrow}

\begin{document}
\preprint{APS/123-QED}

\title{Probing quantum critical phase from neural network wavefunction}

\author{Haoxiang Chen}
 \affiliation{School of Physics, Peking University, Beijing 100871, People's Republic of China.}
 \affiliation{ByteDance Research, Fangheng Fashion Center, No. 27, North 3rd Ring West Road, Haidian District, Beijing 100098, P.R. China}
\author{Weiluo Ren}
 \affiliation{ByteDance Research, Fangheng Fashion Center, No. 27, North 3rd Ring West Road, Haidian District, Beijing 100098, P.R. China}
\author{Xiang Li}
\email{lixiang.62770689@bytedance.com}
 \affiliation{ByteDance Research, Fangheng Fashion Center, No. 27, North 3rd Ring West Road, Haidian District, Beijing 100098, P.R. China}
\author{Ji Chen}
 \email{ji.chen@pku.edu.cn}
 \affiliation{School of Physics, Peking University, Beijing 100871, People's Republic of China.}
 \affiliation{Interdisciplinary Institute of Light-Element Quantum Materials, Frontiers Science Center for Nano-Optoelectronics, Peking University, Beijing 100871, People's Republic of China}

\date{\today}

\begin{abstract}
One-dimensional (1D) systems and models provide a versatile platform for emergent phenomena 
induced by strong electron correlation.
In this work, we extend the newly developed real space neural network quantum Monte Carlo methods to study the quantum phase transition of electronic and magnetic properties.
Hydrogen chains of different interatomic distances are explored systematically with both open and periodic boundary conditions, and fully correlated ground state many-body wavefunction is achieved via unsupervised training of neural networks.
We demonstrate for the first time that neural networks are capable of capturing the quantum critical behavior of Tomonaga-Luttinger liquid (TLL), which is known to dominate 1D quantum systems.
Moreover, we reveal the breakdown of TLL phase and the emergence of a Fermi liquid behavior, evidenced by abrupt changes in the spin structure and the momentum distribution. 
Such behavior is absent in commonly studied 1D lattice models and is likely due to the involvement of high-energy orbitals of hydrogen atoms.
Our work highlights the powerfulness of neural networks for representing complex quantum phases.

\end{abstract}

\maketitle



\section{\label{sec:introduction}Introduction}

Understanding quantum phases emerged and phase transitions in interacting many-electron systems have been an interesting yet challenging question 
in the transdisciplinary field of condensed matter physics and quantum chemistry. 
One-dimensional (1D) interacting systems are believed to have distinct differences from higher-dimensional systems. 
A well-known phenomenon that emerged in 1D is the Tomonaga-Luttinger liquid (TLL), featuring linear gapless modes and quasi-long range correlation.
The TLL behavior applies to a vast range of 1D models, such as the Heisenberg model and the one-band Hubbard model, and stands generally true. 
However, numerical studies have found exceptions with extended terms in model Hamiltonian, showing evidence of magnetic ordering \cite{multi_orb_Hubbard_phase_diag}, or deviation from universality class \cite{other.breakdown-of-TLL}.
These exceptions bring up an important question of whether TLL is universal for more realistic 1D systems.

The hydrogen chain is a 1D system with real atoms, which is a useful platform to probe quantum phases.
Mean-field {\it ab initio} calculations of equispaced hydrogen chains have also revealed magnetic ordering, including ferromagnetic phase, spiral order and various antiferromagnetic orderings \cite{1D_H_chain_doping, H_chain_phase}.
However, the hydrogen chain is one of the emblematic examples of strongly correlated systems,
where quantum fluctuation may be strong enough to destroy mean field approximations.
Therefore, in recent years, amounts of high-level methods have been applied to solve this system
\cite{energy.DMRG,energy.AFQMC,energy.VMC,H_chain_VMC,energy.2RDM}.
These studies have pushed forward the accuracy of ground state energy, leading to more reliable determination of phase transitions,
including dimerization, metal-insulator transition 
and magnetic phase transition \cite{H_chain_VMC,H_chain_phase,Peierls.FCI}.
At large interatomic distance, these advanced calculations have now agreed on a TLL phase with quasi-long range $q=\pi$ antiferromagnetic order, but different behaviors are noted at smaller atomic separation \cite{H_chain_VMC,H_chain_phase}. 
The knowledge of quantum phase transition across a wide range of interatomic distances is yet to converge.
%
%
%

Recent years have witnessed the combination of
deep neural network wavefunction \textit{ansatz} with variational Monte Carlo (VMC)
in solving complex electronic structure \cite{ferminet,nnqmc_excited_states,DeepSolid},
which is referred to as neural network quantum Monte Carlo (NNQMC).
NNQMC achieves accuracy comparable to, or even better than, other methods on the highest level, and provides desirable computational scaling to system size \cite{ferminet,DeepSolid,NNQMC_HEG}.
Besides the ground state energy,
other physical properties have been acquired within the NNQMC framework,
such as polarization \cite{NNQMC_polarization}, wavefunction localization \cite{DeepSolid},
and radial distribution function in boson Helium bulk \cite{NNQMC_He}.
It is now a legitimate question whether NNQMC can reliably probe strong quantum fluctuation and spin correlation, hence describing emergent quantum phase and phase transitions.

In this work, we extend real space NNQMC to exploring quantum phases in hydrogen chain from an \textit{ab initio} perspective.
Calculations are performed over a wide range of interatomic distances,
covering electronic structure from weak to strong correlation.
Real-space neural network wavefunction encodes correlation effects beyond low-energy bands considered in lattice models and second quantized \textit{ab initio} approaches with small basis sets. 
To probe the quantum phases associated with electronic and magnetic structures,
we develop one- and two-body reduced density matrices (1- and 2-RDM) sampling in NNQMC.
Magnetic fluctuation is also tested with the perturbation of an external magnetic field in NNQMC.
In Sec. II, we will describe the key framework of NNQMC and the methodological developments to probe electronic and magnetic structures.
In Sec. III, we present and discuss the main results on spin fluctuation, TLL behavior and the transition to Fermi liquid.
Our results show that neural networks are powerful architectures to represent real space wavefunction of correlated systems, even for capturing quantum critical phases with strong quantum fluctuation, as well as quantum phase transitions.



\section{\label{sec:methods}Methods}
\begin{figure*}
  \centering
  \includegraphics[width=0.9\textwidth]{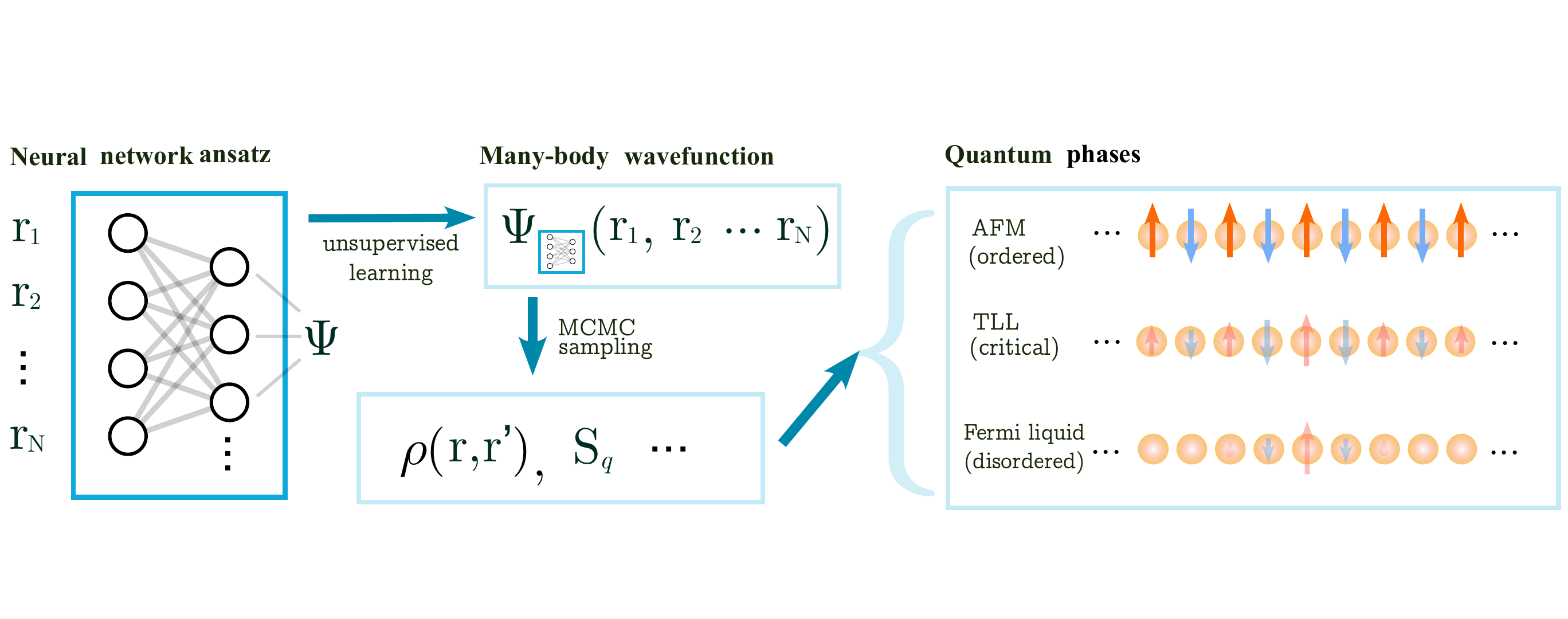}
  \caption{
    \textbf{Workflow of neural network quantum Monte Carlo on hydrogen chain.} 
    The neural network ansatz takes in all electron position, and evaluates the wavefunction value.
    The parameters of the neural network ansatz are optimized to lower the total energy, 
    and the wavefunction is variational.
    With the optimal neural network many-body wavefunction, properties can be sampled with Monte Carlo,
    such as the one- and two-particle reduced density matrices $\rho(r,r'), \rho(r_1,r_2;r_1',r_2')$, 
    and the spin structure factors $S_q$, from which the magnetic order can be probed.
    For an ordered magnetic phase, the magnetism survives to the thermal dynamic limit. 
    Tomonaga-Luttinger liquid (TLL) is critical, where the magnetism is quasi-long ranged.
    Fermi liquid is disordered, and magnetic perturbation induces only local response. 
  }
  \label{fig:overview}
\end{figure*}

The overall workflow of this study is shown in Fig. \ref{fig:overview}.
First, the neural network is trained with an unsupervised learning process according to the variational principle.
After training, electronic and magnetic properties are extracted from the optimal many-body neural network wavefunction, either by direct stochastic samplings or via 1- and 2-RDMs.
The behaviors of different quantum phases are then probed from the calculated properties.

\subsection{\label{sec:method:NNQMC} Neural network quantum Monte Carlo}
Neural network quantum Monte Carlo used in this work is a variant of 
real space variational Monte Carlo (VMC).
In VMC, the solution to the Schrödinger equation is approximated by
trial wavefunction $\Psi_T(r;\alpha)$, known as {\it ansatz}, which is controlled by a set of parameters $\alpha$.
In NNQMC, neural networks are used as {\it ansatz} to enhance the expressivity.
The unsupervised training of neural network in NNQMC proceeds toward obtaining optimized network parameters so that the energy expectation
\begin{equation}
  E_v = \frac{\langle \Psi_T | H | \Psi_T \rangle}{\langle \Psi_T | \Psi_T \rangle}
\end{equation}
takes minimum.
Due to the complexity of ansatz, the integration is carried out with Markov chain Monte Carlo (MCMC).
When the parameters are set, the real space wavefunction is represented by the neural network, 
which takes in electron coordinates and returns the wavefunction value.

In this work, we use FermiNet \cite{ferminet} and DeepSolid \cite{DeepSolid} architectures for open and periodic hydrogen chains, respectively.
For a brief description, the FermiNet ansatz takes a Slater determinant form
\begin{equation}
  \Psi_{\rm net} = {\rm det}[\phi_i(\mathbf{x}_j;{\mathbf{x}_{/j}})]
\end{equation}
to enforce antisymmetry.
$\phi_i$ stands for a neural network orbital, where $\mathbf{x}_j$ is the spin and position of $j$-th electron and $\mathbf{x}_{/j}$ denotes all the spin and position besides the $j$-th electron.
DeepSolid ansatz \cite{DeepSolid} has the form
\begin{equation}
  \Psi_{\rm net} = {\rm det}[{\rm e}^{i \mathbf{k}_i \cdot \mathbf{r}_j} u_{\mathbf{k}_i}(\mathbf{r}_j;\mathbf{r}_{/j})],
\end{equation}
where $u_{\mathbf{k}_i}$ are generalized Bloch wavefunctions, also represented by neural networks.
To encode the translational symmetry, the coordinates $\mathbf{r}_i$ are mapped with periodic functions before going into the neural network.
It is worth noting that $\phi_i$ and $u_{\mathbf{k}_i}$ can capture electron correlations, hence they should be seen differently from single particle orbitals used in the Hartree-Fock method and the Kohn-Sham scheme.

\subsection{\label{sec:method:RDM sampling}Sampling reduced density matrices}
\label{sec:RDM_sampling}
With a real space neural network wavefunction, we can obtain RDMs via implementation of a robust sampling scheme.
1-RDM and 2-RDM are sampled with a $N+1$ or a $N+2$-dimensional integral, where $N$ is the number of electrons, defined as
\setlength{\multlinegap}{2em}
\label{eq:RDM-sampling-no-importance-sampling}
\begin{multline}
  \quad \Gamma^{(1)}_{ij}=   \sum_a \int {\rm d}R \; {\rm d} r_a' \; \
  \Psi^*(R) \; \Psi(R_a') \\
  \times \varphi^*_i(r_a) \varphi_j(r_a') \qquad \qquad \quad
  \label{eq:1-RDM-sampling}
\end{multline}
\begin{multline}
  \Gamma^{(2)}_{ijkl} =  \sum_{a\ne b} \int {\rm d}R \; {\rm d}r_a' \; {\rm d} r_b' \;
  \Psi^*(R) \; \Psi(R_{ab}'') \\
  \times \varphi^*_i(r_a) \varphi^*_j(r_b) \varphi_k(r_a') \varphi_l(r_b')
  ,
  \label{eq:2-RDM-sampling}
\end{multline}
where $R=\{r_1,r_2,\cdots,r_N\}$ contains all electron coordinates, $R_a'=\{r_1,\cdots,r_{a-1},r_a',r_{a+1},\cdots\}$, and
$R_{ab}''=\{r_1,\cdots,r_a',\cdots,r_b',\cdots\}$.
To accelerate the convergence, importance sampling is used and the integrals become,
\begin{align}
  &\Gamma^{(1)}_{ij}
  = \sum_{a} 
  \frac
  {
    \mathbb{E}_{|\Psi|^2,f}\left[
    \frac{\Psi(R_{a}')}{\Psi(R)}
    \frac{\varphi_i^*(r_a) \varphi_j(r_a') }{f(r_a')}  
    \right]
  }
  {N_i N_j} \\ 
  &\Gamma^{(2)}_{ijkl}
  = \sum_{a\ne b} 
  \frac
  {
    \mathbb{E}_{|\Psi|^2,f}\left[
    \frac{\Psi(R_{ab}'')}{\Psi(R)}
    \frac{\varphi_i^*(r_a) \varphi_j^*(r_b) \varphi_k(r_a') \varphi_l(r_b')}{f(r_a')f(r_b')}  
    \right]
  }
  {N_i N_j N_k N_l}
  \label{eq:RDM-sampling}
\end{align}
where $\mathbb{E}_{|\Psi|^2,f}$ stands for the average according to sampling of $R$ and $r_a',r_b'$ according to $|\Psi|^2$ and $f(r')=\sum_i |\varphi_i (r')|^2$, respectively.
$N_i= \mathbb{E}_f \left[{|\varphi_i(r_a')|^2/f(r_a')} \right]^{1/2}$ is defined to normalize $\varphi_i$.

For periodic systems, while wavefunction $\Psi$ is periodic, localized orbitals $\varphi_i$ are not.
Hence, RDMs for periodic systems are sampled in a slightly different way, for example 1-RDM is evaluated as
\begin{multline}
  \Gamma^{(1)}_{ij}
  =  \sum_a \int_{\rm cell} {\rm d}R {\rm d} r_a' \Psi^*(R) \; \Psi(R_a') \\
  \times \sum_{T=-\infty}^{\infty}\varphi^*_i(r_a+aT) \sum_{T=-\infty}^{\infty}\varphi_j(r_a'+aT) \
  , 
\end{multline}
where $a$ is lattice constant, and the summation over $T$ counts for periodicity of $\Psi$ and is shifted to a single primitive cell.
The evaluation of orbital $i$ should in principle contain all the periodic images,
and we take a real space cutoff to include images up to the next nearest.

From RDMs, the spin order parameter and spin-spin correlation function can be calculated.
$\bra S_i^z \ket $ is calculated with the diagonal elements of 1-RDM.
\begin{equation}
  \bra S_i^z S_j^z \ket = 
  \Gamma^{(1)\ua}_{ii} + \Gamma^{(1)\da}_{ii} 
  + \Gamma^{(2)\ua\ua}_{ijij} + \Gamma^{(2)\da\da}_{ijij}
  - \Gamma^{(2)\ua\da}_{ijij} - \Gamma^{(2)\ua\da}_{jiji},
  \label{eq.appx:spin-structure}
\end{equation}
where $\Gamma^{(\mu)}_{ij}=\bra c_{i\mu}^\dagger c_{j\mu} \ket$, 
$\Gamma^{(2)\mu\nu}_{ijkl}=\bra c_{i\mu}^\dagger c_{j\nu}^\dagger c_{l\nu} c_{k\mu} \ket$.
The electron occupation on a certain basis set is calculated as $n={\rm Tr}(\mathbf{\Gamma}^{(1)} \mathbf{S})$,
where $S$ is the overlap matrices, and $\Gamma^{(1)} = \Gamma^{(1)\ua} + \Gamma^{(1)\da}$.
Natural orbital (NO) is the eigenvector of 1-RDM, 
and the corresponding eigenvalue gives the NO occupation.
Wannier orbitals are acquired by the eigenvector of the position operator in the space of the $N$ (the number of electrons) most occupied NOs.

\subsection{\label{sec:medthod:structure factor}Sampling structure factor and momentum distribution}
The charge and spin structure factors are calculated by real space MCMC sampling.
By performing Fourier transformation of $r_j,r_k$ separately,
the charge structure factor is,
\begin{equation}
  S_{\rho}(q)= \bra (n_{q\uparrow} + n_{q\downarrow})(n^*_{q\uparrow} + n^*_{q\downarrow}) \ket_c,
\end{equation}
and the spin structure factor is,
\begin{equation}
  S_{\sigma}(q)= \bra (n_{q\uparrow} - n_{q\downarrow})(n^*_{q\uparrow} - n^*_{q\downarrow}) \ket_c,
  \label{eq:spin-structure-factor}
\end{equation}
where the subscript $c$ denotes the connect part, and
$n_{q\uparrow}=\int{{\rm d}r\;{\rm e}^{iqr}n_{\uparrow}(r)}$.
For each $q$ point, $n_{q\uparrow}$ can be efficiently sampled
by $n_{q\uparrow}=\mathbb{E}_{i\in {\rm spin-up}} [ {\rm e}^{i q r_i }] $, 
where $r_i$ is the position of all spin-up electrons in each MCMC step.

The momentum distribution function is defined as $ n(k)\equiv\bra c_k^\dagger c_k\ket $.
By Fourier transformation into real space, 
\begin{align}
  n(k) &= \bra \int \mathrm{d}r \mathrm{d}r' \mathrm{e}^{ikr} c^\dagger(r) \mathrm{e}^{-ikr'} c(r') \ket \\ 
  &= \int \mathrm{d}r' \mathrm{e}^{ikr'} \int \mathrm{d}r \bra c^\dagger(r+r') c(r) \ket 
\end{align}
where $\bra c^\dagger(r+r') c(r)\ket $ can be sampled from real space.
A set of electron positions is sampled with importance,
\begin{equation}
  \int \mathrm{d}r \bra c^\dagger(r+r') c(r) \ket = \mathbb{E}_{|\Psi|^2} \left[ \frac{\Psi(r_1+r',r_2,...)}{\Psi(r_1,r_2,...)} \right]
\end{equation}
where ${r_1,\cdots,r_N}$ are generated from MCMC with weight $|\Psi|^2$.
$r'=ma$, $m\in \{0,1,\cdots,N-1\}$ and $a$ is lattice constant.

For small periodic systems, we use the twisted boundary condition, reading $c^\dagger(r+L)=\mathrm{e}^{i\theta}c^\dagger(r)$.
The equations remain unchanged, except the $k$-points are shifted to $k=(2m\pi+\theta)/N$, 
where $N$ is the number of atoms in the supercell, and $m$ is also an integer less than $N$.


\section{\label{sec:results}Results and Discussion}
\subsection{Validation of neural network wavefunctions}

\begin{figure}[ht]
  \centering
  \includegraphics[width=\columnwidth]{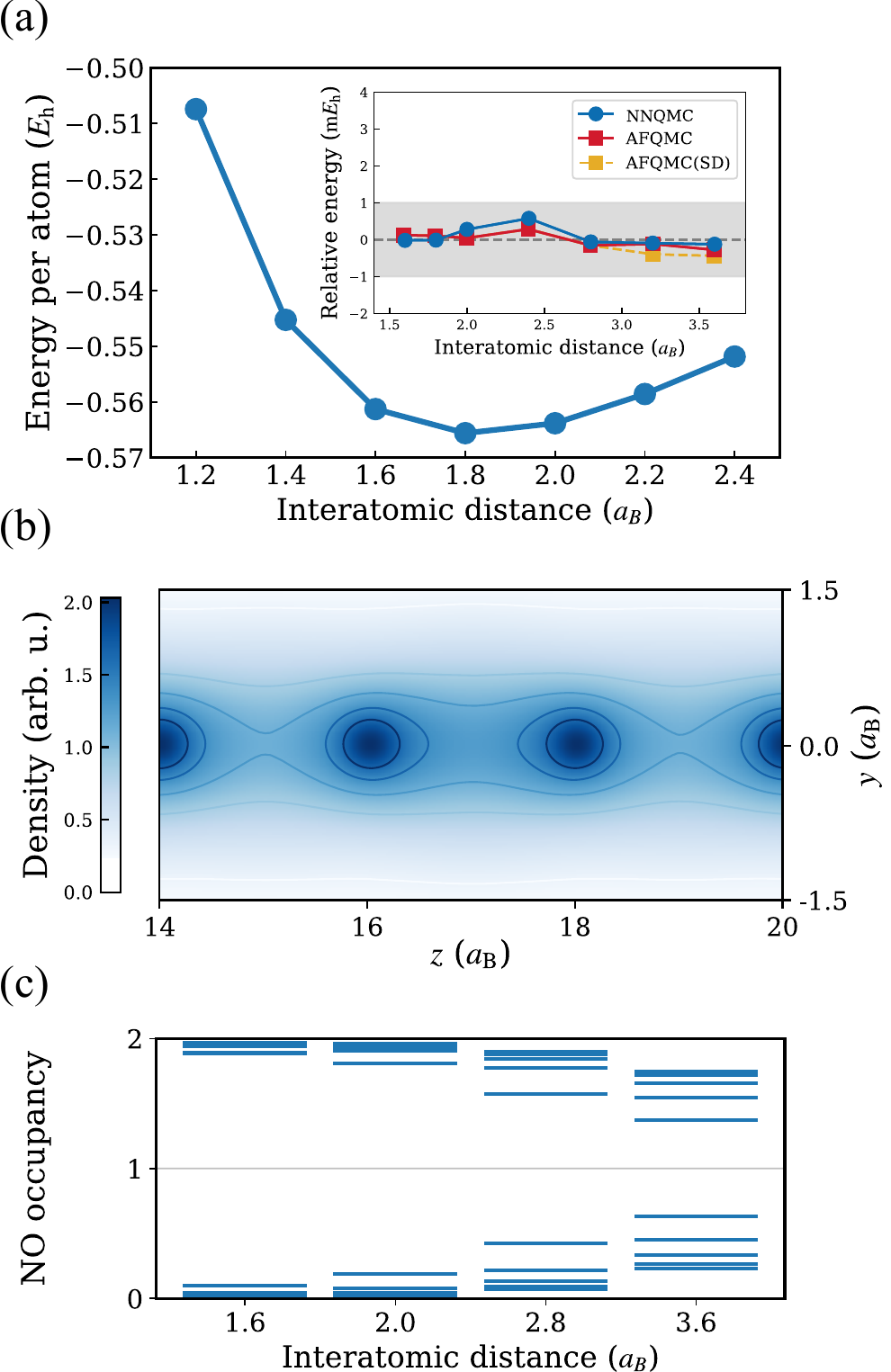}
  \caption{\textbf{Validation of neural network wavefunctions.} 
  (a) Potential energy curve for hydrogen chain with $N$=16 sites at different interatomic distances. 
  The inset shows a comparison of a $N$=10 chain with other high-level theories reported in ref. \cite{H_chain_energy}, including auxiliary field quantum Monte Carlo (AFQMC), and benchmarked by multi-reference configuration interaction with quadruples (MRCI+Q), where results were extrapolated to the complete basis set limit.
  The gray shade indicates the chemical accuracy of $1 \;{\rm m}E_{\rm h}$.
  Statistic errors in NNQMC are negligible, less than $10^{-4}\; E_{\rm h}$ for total energy.
  (b) The electron density distribution plotted on the plane passing through the chain (OBC, $N=16$, $R=2.0\; a_{\rm B}$), showing a case of dimerization.
  (c) Natural orbital (NO) occupation. 
  Each horizontal line represents the electron number on a NO, obtained from the diagonalization of 1-RDMs.
  Coulomb correlation drives the NO occupation away from 0 or 2 at large interatomic distance.
  }
  \label{fig:verification}
\end{figure}

First of all, to demonstrate the accuracy of NNQMC, 
we compare the energy of few-atom equispaced hydrogen chains under the open boundary condition (OBC) 
with the other state-of-the-art methods (Fig. \ref{fig:verification}).
In our calculations, the equilibrium bond length is around $1.8\; a_{\rm B}$, and the energy minimum is $0.56561(5)\;{\rm E_h}$, as shown in Fig. \ref{fig:verification}(a).
The inset shows a comparison of the energy of a 10-atom chain with other high-level theories reported in the literature. 
One can see that for these short chains, NNQMC and other high-level theories can reach satisfactory consistency, within $1 \; {\rm mE_h}$ for energy calculation. 
In these other calculations, a limited basis set is used and an extrapolation to the complete basis set limit is necessary to achieve good energy evaluation \cite{H_chain_energy}.
In our NNQMC calculations, real space wavefunction is achieved, which automatically corresponds to the complete basis set limit.

Overall, the energy agreement indicates a reliable ground state wavefunction obtained by NNQMC, and it provides confidence to further analyze wavefunction to extract physical properties and to extend to larger systems.
Here, we first revisit the well-known Peiers instability of hydrogen chains, 
where electrons tend to form pairs at neighboring sites to lower the total energy \cite{Peierls.H-ladder-mol-atm,Peierls.H2-chain-mol-atm}.
The electron density distribution plotted in Fig. \ref{fig:verification}(b) shows a typical dimerization feature.
The minima of charge density appear at the middle points between two proton sites,
and the minima show a periodicity of two sites.

It is also a common insight that chains with large interatomic distances have stronger correlation effects.
The strong correlation nature can be shown by natural orbitals,
in which the 1-RDM is diagonal.
At small interatomic distances, electrons can easily hop between neighbor atoms, making the kinetic energy more important, hence the NOs are either doubly occupied or empty.
As the interatomic distance grows, NOs occupations are driven towards 1, as electrons tend to localize around nuclei.
Coulomb repulsion dominates, and it prevents two electrons from occupying the same orbital.
Fig. \ref{fig:verification} (c) presents the electron occupation number on each NO, obtained with 1-RDM re-sampled from real space neural network wavefunctions.
It indeed displays the expected strong correlation nature of hydrogen chains,
which would be underestimated by mean-field theories.

\subsection{Tomonaga-Luttinger liquid\label{sec:AFM}}
  \begin{figure*}
    \centering
    \includegraphics [width=\textwidth]{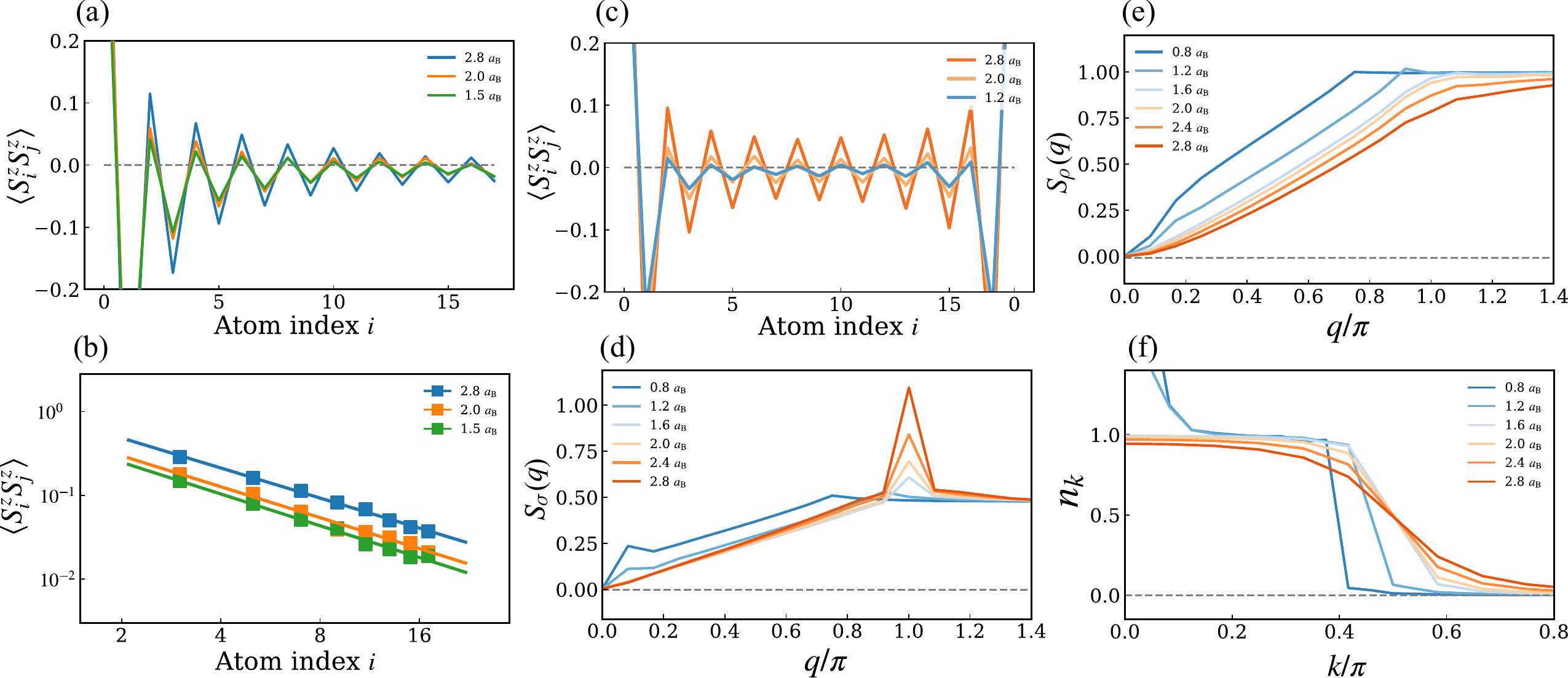}
    \caption{\textbf{Tomonaga-Luttinger liquid behavior of hydrogen chain.}
    (a) Spin-spin correlation functions calculated with chains of $N=18$ under OBC. 
    (b) Critical exponent fitted from the spin-spin correlation functions in panel (a).
    (c-f) Results calculated with $N=24$ under PBC.
    (c) Spin-spin correlation function.
    (d) Spin structure factor.
    (e) Structure factor.
    (f) Momentum distribution.
    }
    \label{fig:TLL}
  \end{figure*}

Having verified the reliability of NNQMC calculations,
we further investigate the spin degree of freedom.
We first focus on large-distance chains, 
in which Coulomb energy is much greater than kinetic energy,
and electrons are strongly correlated.
These systems are expected to behave similarly to the one-band Hubbard model,
and should host antiferromagnetic fluctuation. 
Spin fluctuation is depicted by spin-spin correlation $G_{\rm ss}^{z}(i-j)\equiv\langle S_i^z S_j^z\rangle$.
In analog to sites in the second quantization scheme,
here sites are defined with localized atomic orbitals.
RDMs are sampled with localized orbitals, as shown in \ref{sec:method:RDM sampling},
based on which the spin-spin correlation function is calculated.

Fig. \ref{fig:TLL}(a) shows the open chain results for spin-spin correlation between the left-most and the $i$-th site, which goes to zero at long separation as a result of quantum fluctuation.
The correlation function alters signs for even and odd sites,
indicating that AFM fluctuation plays a major role.
A similar result is also shown in Fig. \ref{fig:TLL}(c), where the periodic boundary condition (PBC) is employed.
The choice of OBC \cite{Peierls.FCI} or PBC \cite{H_chain_VMC} affects convergence to system size, but both calculations can capture spin fluctuation.
Besides the correlation function, we also implement a direct magnetic perturbation to probe the spin fluctuation, as described in SI Sec. \uppercase\expandafter{\romannumeral2},
which confirms the AFM fluctuation.

Moreover, AFM order of one-dimensional system is expected to be quasi-long-range,
i.e. the correlation is expected to decay algebraically with respect to the site separation for large enough separation.
The conformal field theory suggests for one-dimensional Heisenberg model,
spin-spin correlation decays as
\begin{equation}
  \label{eq:critical_exponent_fitting}
  G_{\rm ss}^{z} (i) \propto (-)^{i} \;  {i}^{-\eta} \ln{i},
\end{equation}
where $\eta$ is the critical exponent and is predicted to be $\eta=1$,
and the logarithm correction comes from finite size effect \cite{CFT_critical_exponent1}.
We expect that the hydrogen chain in a strong correlation regime belongs to this universality class.
Fig. \ref{fig:TLL}(b) shows a fit to the decay of the correlation function to determine the critical exponent.
For the $2.8 \; a_{\rm B}$ chain, 
the critical exponent is found to be 1.16(5), close to $\eta=1.11(1)$ reported in \cite{H_chain_phase}. 
The critical exponent $\eta$ stays unchanged for medium to large distances,
indicating the absence of phase transitions to other AFM orders in this range.
The deviation of $\eta$ from exact 1 is a result of finite size effect.
It is worth noting that the correlation amplitude is smaller for the even site than the odd site, which was also observed in previous DMRG calculations \cite{H_chain_phase}.
Nevertheless, the even-odd difference does not affect the critical exponent.
When fitting with even or odd sites separately,
the results are 1.2(1) and 1.18(8), respectively.

For PBC chains, the quasi-long range spin-spin correlation is shown with
spin structure factor, in Fig. \ref{fig:TLL}(d).
The spin structure factor is defined by eq. \ref{eq:spin-structure-factor},
which is the Fourier series of spin-spin correlation.
For chains with bond lengths larger than 1.6 $a_{\rm B}$,
spin structure factor shows a single peak at $q=\pi$,
showing AFM correlation dominates.
At the limit of $q\rightarrow 0$, 
the spin structure factor disperses linearly, 
which is a signature of TLL.
The slope of the spin structure factor is related to the central charge $K_\sigma$ 
by $ \lim_{q\rightarrow 0} S_{\sigma}(q)/q = K_\sigma /\pi $,
which is essential to gapless spin excitation \cite{other.linear_structure_factor}.
Here, for the $2.8\;a_\mathrm{B}$ chain, 
$K_\sigma$ is determined to be $1.067$, 
which is very close to the theoretical value $K_\sigma=1$.
For chains with distance larger than $1.6\;a_\mathrm{B}$ (warm colors in Fig. \ref{fig:TLL}(d)),
$K_\sigma$ takes similar values.
These evidences confirm the existence of the gapless linear spin excitation mode in the hydrogen chain.

Excitations in TLL are collective modes, but the spin and charge are carried by separate modes. 
While the spin channel is gapless, 
the charge excitation is gapped.
The charge structure factors are shown in Fig. \ref{fig:TLL}(e).
One can notice that a quadratic behavior occurs at small wave vector $q$ for large distance chains,
indicating a gap in the charge mode.
The momentum distribution $n(k)$ in Fig. \ref{fig:TLL}(f) also shows also a sign of TLL for large-distance chains, i.e. 
the momentum distribution changes continuously at $k=\pi/2$. 
In SI Sec. \uppercase\expandafter{\romannumeral1} C
we verify that this is not an artifact of finite size.
To sum up, we demonstrate that the neural network wavefunctions are powerful enough to capture the TLL quantum critical phase,
even without specific designs of the neural network structure.


\subsection{\label{sec:self-doping}Transition to Fermi liquid}
For chains with interatomic distances less than $1.6\;a_\mathrm{B}$, shown by cold colors in Fig. \ref{fig:TLL}(d-f), very different behaviors are observed, indicating a transition to a different quantum phase.
Hereby, we further examine the nature of this transition.

For the spin structure factor in Fig. \ref{fig:TLL}(d),
the major peak shifts away from $q=\pi$,
and another kink appears at a small wave length.
With decreasing interatomic distance,
the major peak further shifts away from $\pi$.
Similarly, a kink also rises at the same $q$ in the charge structure factor shown in Fig. \ref{fig:TLL}(e).
For the momentum distribution (Fig. \ref{fig:TLL}(d)), 
the transition near $k=\pi/2$ is much sharper when the interatomic distance is reduced, indicating a non-zero quasi-particle weight.
In addition, for chains with bond lengths smaller than $1.6\;a_{\rm B}$,
a sharp peak develops near $k=0$, and $n(k)$ is even greater than one.
These features in momentum distribution differs from TLL strongly and hence indicates a transition to the Fermi liquid at small interatomic distances.

\begin{figure*}[ht]
  \centering
  \includegraphics[width=\textwidth]{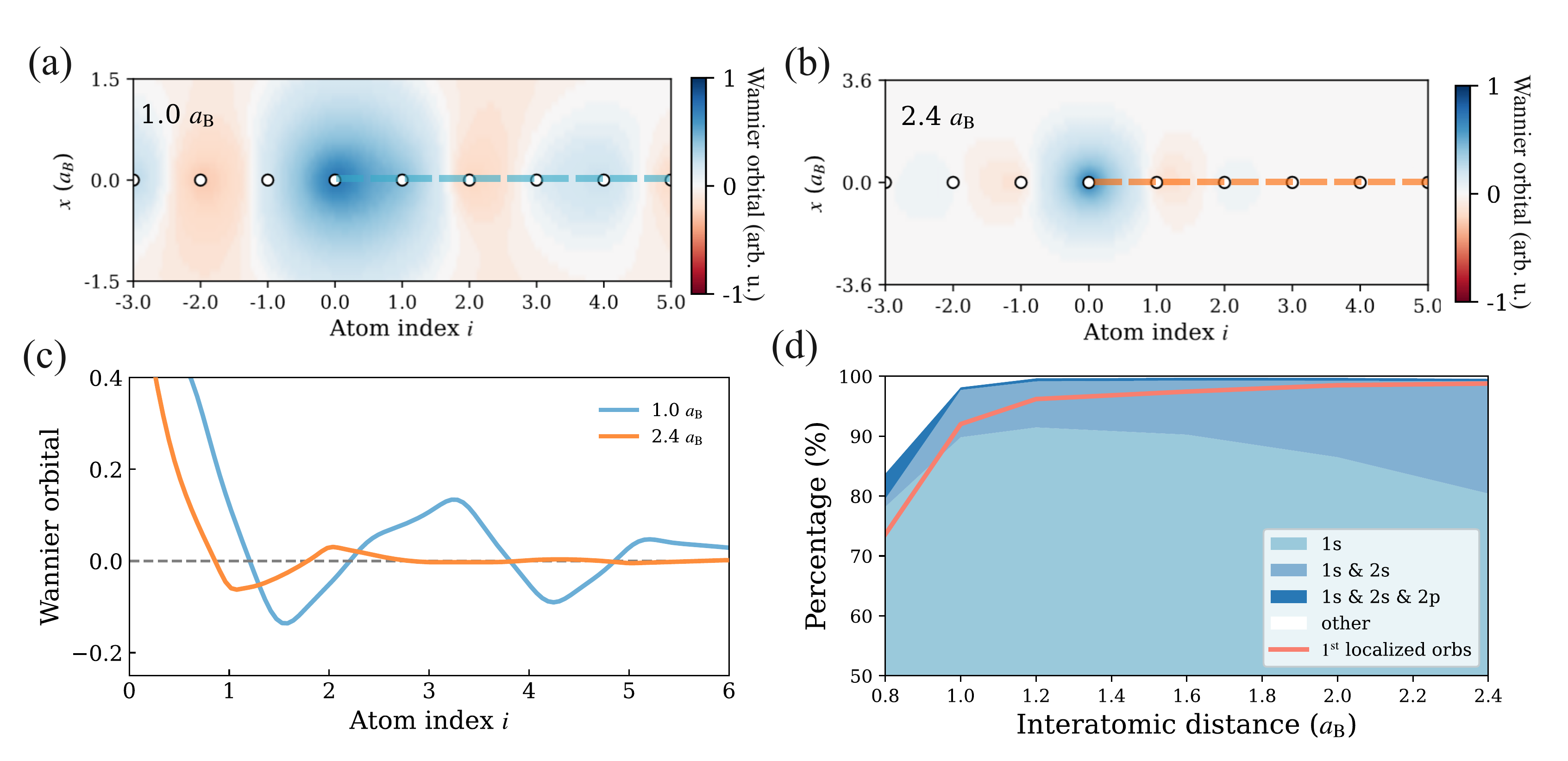}
  \caption{\textbf{Transition due to self-doping mechanism.}
  (a-b) Wannier orbitals for (a) a band conductor chain with interatomic distance $1.0\;a_{\rm B}$ and (b) a Mott insulator chain with interatomic distance $2.4\;a_{\rm B}$.
  The blue (red) shape shows the regions where the Wannier orbital is positive (negative).
  (c) The comparison of Wannier orbital value for chains in (a) and (b) along the chain axis.
  (d) Electron occupation number on atomic orbitals. 
  The shades show accumulated electron number on $1s$, $2s$ and $2p$ orbitals,
  as percentage of the total electron number.
  The orange line shows the number on the lowest localized orbitals, 
  which are linear combinations of atomic orbitals.
  }
  \label{fig:self-doping}
\end{figure*}

The physical insight behind the transition is that as the interatomic distance becomes comparable to the size of the hydrogen atom,
electrons become diffusive in real space.
In these scenarios, the "self-doping" mechanism \cite{H_chain_phase} kicks in.
The conventional understanding of the hydrogen chain is based on the one-band models, where only the hydrogen $1s$ orbital is considered \cite{Hubbard_critical_exponent, CFT_critical_exponent1,CFT_critical_exponent2}.
The one-band models can well capture the physics of large-distance chains as discussed in the previous section.
However, as the electrons become more diffusive, they start to occupy bands formed by orbitals higher than $1s$.
In the band theory view of the self-doping mechanism,
Fermi surface cuts through multiple bands,
and electrons are self-doped from the $1s$ band to higher ones.
The self-doping mechanism is responsible for the Fermi liquid behavior observed in Fig. \ref{fig:TLL}.
At the same time, one can expect a transition from the Mott insulator to the band conductor to occur when reducing the interatomic distance \cite{H_chain_VMC,H_chain_phase}.
In Fig. \ref{fig:self-doping}(a-b), we present the Wannier orbitals acquired neural network wavefunctions, which further support the transition.

To demonstrate the self-doping mechanism explicitly,
we further check the electron occupation number on a set of atomic orbitals. In Fig. \ref{fig:self-doping}(c), we first project the real-space neural network wavefunction onto the 6-31G** basis set.
From medium to strong correlation regime,
all the electrons dwell on the $1s$ and $2s$-like orbitals of this basis set.
Meanwhile, for separation less than 1.0 $a_{\rm B}$, 
a non-negligible amount of electrons occupy orbitals other than the lowest $1s$, $2s$ and $2p$-like orbitals.
To exclude the effect of the choice of basis set, 
we also combine basis functions into localized orbitals using the meta-L\"{o}wdin scheme and then calculate the 
occupation of the first localized orbital from RDMs.
As shown by the orange curve in Fig. \ref{fig:self-doping},
nearly all electrons distribute on the first localized orbitals for medium to large atomic distance.
However, when interatomic distance decreases, the occupation of the first localized orbitals decreases very fast. 
For atomic distances smaller than 1.0 $a_{\rm B}$,
only less than 90\% electron distributes on the first localized orbitals.

From the above analyses, it is now clear why the ground state physics of large-distance hydrogen chains are similar to the one-band models.
This is because electrons tend to localize at each site when hydrogen atoms are separated with a large distance, and hence the ground state can be properly described by a minimal basis set \cite{energy.2RDM,Peierls.FCI}.
Consequently, the one-band Hubbard model \cite{Hubbard_critical_exponent} is valid, alternatively Heisenberg model \cite{CFT_critical_exponent1,CFT_critical_exponent2} can be applied to the strong correlation limit.
With both models, the TLL behavior is predicted.
The complexity rises with more realistic interactions in the hydrogen chain.
When hydrogen atoms are close, self-doping occurs, necessitating multi-band models for understanding the low-energy physics.
At the same time, as the orbitals become diffusive, the interaction naturally extends beyond the nearest sites, leading to the breakdown of TLL.
With NNQMC, we can now solve hydrogen chains in real space, breaking through the limitation of small basis sets and shedding light on which higher orbitals and farther atomic sites are involved in the quantum phases.  
In other theoretical works, it was also predicted that TLL can survive with multiple Fermi points \cite{other.t1-t2-Hubbard-TLL.compt,other.t1-t2-Hubbard-TLL.theory} but interaction extending beyond nearest sites may lead to the breakdown of TLL \cite{other.breakdown-of-TLL}.

\section{\label{sec:discussion}Conclusion}

To conclude, {\it real space} NNQMC is a powerful method to study quantum critical phases and phase transitions in {\it ab initio} systems.
In this work, we focus on the hydrogen chain, where the interatomic distance can be tuned to generate a sequence of 1D systems from weak to strong correlation regime.
The charge and spin structures are computed with the implementation of reduced density matrices under both open and periodic boundary conditions.
The trained neural network wavefunctions not only capture the correct TLL behavior at large-distance regime but also predicts a transition to the Fermi liquid phase when the interatomic distance between hydrogen is smaller than $1.6\;a_{\rm B}$.
The self-doping mechanism of the transition is captured by the real-space wavefunctions, which automatically encodes the correlation effects of high-energy orbitals.

Moreover, it is worth emphasizing that NNQMC has been rising as one of the most powerful {\it ab initio} approaches in solving real molecules and materials \cite{review_on_NNQMC, review_on_NNQMC_PBC}. 
In particular, real space NNQMC can achieve better accuracy and efficiency than state-of-the-art quantum chemistry methods for energy calculations.
The methodological developments of this work, mainly the combining of the real space wavefunction with the second quantization representation through RDM samplings, can further extend the application of NNQMC.
This combination allows us to gain insights into the electronic structure of the systems and related effective models.
Also, it provides a general route to obtain physical quantities such as the structure factors and momentum distribution reported in this work, and potentially other observables such as the spectrum density \cite{RDM_spectrum}. 
Besides, the same methodology is compatible with transition RDMs, 
which can be related to photoemission.
Last but not least, all these can be achieved within real space, and the basis set incompleteness error is hence circumvented.  
Therefore, we can envision a broad range of applications of our methods to study quantum phase transitions and chemical transitions.

\begin{acknowledgments}
This work was supported by the National Key R\&D Program of China under
Grant No. 2021YFA1400500, the Strategic Priority Research Program of the Chinese Academy of Sciences under Grant No. XDB33000000, and National Science Foundation of China under Grant No. 12334003. We are grateful for computational resources provided by the High Performance Computing Platform of Peking University.
\end{acknowledgments}

\bibliography{apssamp}
\bibliographystyle{unsrt}

\end{document}